\documentclass[fleqn,usenatbib]{mnras}

\usepackage{newtxtext,newtxmath}

\usepackage[T1]{fontenc}

\DeclareRobustCommand{\VAN}[3]{#2}
\let\VANthebibliography\thebibliography
\def\thebibliography{\DeclareRobustCommand{\VAN}[3]{##3}\VANthebibliography}

\usepackage{graphicx}	
\usepackage{amsmath}	

\title[Precise NIR photometry, accounting for PWV at SSO]{Precise near-infrared photometry, accounting for precipitable water vapour at SPECULOOS Southern Observatory}

\author[P. P. Pedersen et al.]{Peter P. Pedersen$^{1}$\thanks{E-mail: ppp25@cam.ac.uk}, 
C. A. Murray$^{1,2}$, 
D. Queloz$^{1}$,
M. Gillon$^{3}$,
B. O. Demory$^{4}$,
\newauthor
A. H. M. J. Triaud$^{5}$,
J. de Wit$^{6}$,
L. Delrez$^{3,7}$,
G. Dransfield$^{5}$,
\newauthor
E. Ducrot$^{8}$\thanks{Paris Region Fellow, Marie Sklodowska-Curie Action}, 
L. J. Garcia$^{3}$, 
Y. G\'omez Maqueo Chew$^{9}$,
M. N. G{\"u}nther$^{12}$\thanks{ESA Research Fellow}, 
\newauthor
E. Jehin$^{7}$,
J. McCormac$^{11}$,
P. Niraula$^{6}$,
F. J. Pozuelos$^{3,7}$,
B. V. Rackham$^{6,10}$\thanks{51 Pegasi b Fellow}, 
\newauthor
N. Schanche$^{4}$,
D. Sebastian$^{5}$, 
S. J. Thompson$^{1}$, 
M. Timmermans$^{3}$,
R. Wells$^{4}$
\\
$^{1}$Cavendish Laboratory, JJ Thomson Avenue, Cambridge CB3 0HE, UK\\
$^{2}$Department of Astrophysical and Planetary Sciences, University of Colorado Boulder, 2000 Colorado Ave, Boulder, CO 80309, USA\\
$^{3}$Astrobiology Research Unit, University of Li\`ege, All\'ee du 6 ao\^ut, 19, 4000 Li\`ege (Sart-Timan), Belgium \\
$^{4}$University of Bern, Center for Space and Habitability, Gesellschaftsstrasse 6, 3012 Bern, Switzerland \\
$^{5}$School of Physics and Astronomy, University of Birmingham, Edgbaston, Birmingham B15 2TT, United Kingdom \\
$^{6}$Department of Earth, Atmospheric and Planetary Sciences, MIT, 77 Massachusetts Avenue, Cambridge, MA 02139, USA \\
$^{7}$Space sciences, Technologies and Astrophysics Research (STAR) Institute, University of Li\`ege, Belgium\\
$^{8}$AIM, CEA, CNRS, Universit\'e Paris-Saclay, Universit\'e de Paris, F-91191 Gif-sur-Yvette, France\\
$^{9}$Instituto de Astronom\'ia, Universidad Nacional Aut\'onoma de M\'exico, Ciudad Universitaria, Ciudad de M\'exico, 04510, M\'exico \\
$^{10}$Department of Physics, and Kavli Institute for Astrophysics and Space Research, Massachusetts Institute of Technology, Cambridge, MA 02139, USA\\
$^{11}$Department of Physics, University of Warwick, Coventry, CV4 7AL, UK\\
$^{12}$European Space Agency (ESA), European Space Research and Technology Centre (ESTEC), Keplerlaan 1, 2201 AZ Noordwijk, The Netherlands
}

\date{Accepted XXX. Received YYY; in original form ZZZ}

\pubyear{2022}

\begin{document}
\label{firstpage}
\pagerange{\pageref{firstpage}--\pageref{lastpage}}
\maketitle

\begin{abstract}
The variability induced by precipitable water vapour (PWV) can heavily affect the accuracy of time-series photometric measurements gathered from the ground, especially in the near-infrared. We present here a novel method of modelling and mitigating this variability, as well as open-sourcing the developed tool -- \textsf{Umbrella}. In this study, we evaluate the extent to which the photometry in three common bandpasses (\textit{r’}, \textit{i’}, \textit{z’}), and SPECULOOS’ primary bandpass (\textit{I+z’}), are photometrically affected by PWV variability. In this selection of bandpasses, the \textit{I+z’} bandpass was found to be most sensitive to PWV variability, followed by \textit{z’}, \textit{i’}, and \textit{r’}. The correction was evaluated on global light curves of nearby late M- and L-type stars observed by SPECULOOS' Southern Observatory (SSO) with the \textit{I+z’} bandpass, using PWV measurements from the LHATPRO and local temperature/humidity sensors. A median reduction in RMS of 1.1\% was observed for variability shorter than the expected transit duration for SSO's targets. On timescales longer than the expected transit duration, where long-term variability may be induced, a median reduction in RMS of 53.8\% was observed for the same method of correction.
\end{abstract}




\section{Introduction} \label{sec:intro}

Ground-based photometric observations are affected by atmospheric variability. A major source of this contamination comes from the multitude of molecular absorption lines which are known to affect atmospheric transmission. Predominantly in the near-infrared, time-varying amounts of H$_2$O in different layers of the atmosphere affect ground-based observations across a wide range of wavelengths -- with the amount of H$_2$O in a column of our atmosphere quantified as the amount of `precipitable water vapour' (PWV), normally quoted in millimetres. O$_3$, O$_2$, CO$_2$ and CH$_4$, amongst other molecular absorption lines, likewise play a role, however, often to a lesser extent \citep{smette2015molecfit}. 

To mitigate the majority of atmospheric effects, techniques such as differential photometry are commonly adopted \citep{howell2006handbook}. They involve simultaneous observations of multiple objects in a field of view to estimate first-order changes of atmospheric transmission and instrumental effects over the course of an observational period. In differential photometry, objects of similar brightness and spectral energy distribution are used to calculate an ‘artificial’ comparison star. For example, in \citet{murray2020photometry}, the artificial comparison star was created by applying a weight to all the objects in the field of view, accounting for their effective temperature, noise, variability and distance to the target of interest. However, depending on the observational bandpass and an object's spectral energy distribution, the net flux observed on the ground can be seen to vary from object to object if the atmosphere is time-variable in its composition -- leading to a second-order differential effect which cannot be corrected by differential photometry.

This effect is a particularly significant problem for high-precision ground-based photometric studies of cool stars, such as M dwarfs and later types \citep[e.g.,][]{blake2008near, irwin2008mearth, tamburo2022perkins}. This is due to the large disparity in spectral energy distribution frequently observed in a field of view, as cool stars are typically much redder than the comparison stars suitable for differential photometry in any given field of view. To reduce this effect, one could observe with a narrow-band filter \citep[e.g.][]{10.1117/12.2561467}, but at the cost of instrumental precision due to the reduction in photons collected. Instead, most of these studies have relied on post-correction methods to reduce the nature of this effect, with varying degrees of success.

\citet{irwin2011angular} developed a ``common mode'' approach, which they applied to MEarth's northern survey (715\,nm long-pass filter). This approach involved using median values over 30-minute periods from multiple simultaneous observations of similar type objects, from 8 independent telescopes. The common mode, scaled per target via a least-squares optimisation, was then used to correct for atmospheric effects over the course of a target's observation run. They noted that the scaling values correlated with stellar type.

Some observational bandpasses can be extremely sensitive to PWV changes, such as MEarth's bandpass and the primary bandpass used in this study (\textit{I+z'}, 700 -- 1100~nm). Such bandpasses induce false variability, including structures able to mimic a transit feature or to hide a real one in differentially resolved light curves when subject to PWV changes over the course of an observation. To correct for this variability, atmospheric transmission profiles in the near-infrared can be modelled with tools such as Molecfit \citep{smette2015molecfit}. These models can be used to correct photometric observations when the spectra of the observed objects are also known. Methods of quantifying PWV changes are therefore necessary for photometric surveys which use sensitive bandpasses, with temporal resolutions at least half the minimum expected transit duration to resolve such features, to follow Nyquist sampling.

In astronomy, on-site atmospheric transmission profiles are often inferred by spectrographs or by multi-band photometers with strategically placed narrow-band filters. The aTmcam multi-band instrument \citep{li2012atmcam,li2014monitoring} for instance, located on Cerro Tololo at $\sim$2200~m, was able to quantify atmospheric PWV with a stated precision $\sim$0.6~mm. A similar instrument called CAMAL \citep{baker2017monitoring}, located on Mount Hopkins at $\sim$2600~m (same site as MEarth's northern facility), had a stated precision of better than 0.5~mm in dry conditions (PWV < 4~mm). The use of a spectrograph on the other hand, such as in \citet{li2017temporal}, gained a precision of 0.11~mm when evaluating high-resolution near-infrared H-band spectra of hot stars from the APOGEE spectrograph (located on the Apache Point Observatory at $\sim$2800~m), calibrated with GPS-derived PWV values.

PWV values derived from timing delays in GPS signals have historically been used for meteorological studies, with large networks of GPS-PWV derived data in the public domain, such as the SuomiNet project \citep{ware2000suominet} -- typically with a precision of $\sim$1~mm. Remote-sensing satellites have likewise enabled wide spatial and temporal coverage of a multitude of atmospheric parameters globally. \citet{marin2015estimating} were able to estimate PWV in very dry conditions at the Chajnantor plateau (at $\sim$5100~m) with historical observations made by the now-decommissioned GOES-12 satellite and validated by an on-site radiometer. They attained absolute relative errors of 51\% and 33\% over the ranges 0~--~0.4~mm and 0.4~--~1.2~mm, respectively. Similar work was achieved in \citet{valdes2021monitoring}, yielding better uncertainties for Cerro Paranal at around 27\%, validated similarly with an on-site radiometer.

Radiometers derive PWV values from water vapour emission lines in the GHz region. At high altitudes, or low PWVs, the 183~GHz emission line is often observed. At the Paranal Observatory (with an elevation of $\sim$2600~m), a 183~GHz based radiometer was commissioned in October 2011, the Low Humidity and Temperature Profiling microwave radiometer \citep[LHATPRO;][]{kerber2012water}, located on the VLT platform. It has a quoted accuracy of better than 0.1~mm (when PWV between 0.5~--~9 mm) and a precision of 0.03~mm, with a saturation limit of 20~mm.

The motivation for this study derives from the SPECULOOS \citep[Search for habitable Planets EClipsing ULtra-cOOl Stars;][]{gillon2018searching, delrez2018speculoos, murray2020photometry, 2020arXiv201102069S, 10.1117/12.2563563} project, a ground-based photometric survey targeting nearby (<~40~pc) late M- and L-type stars, with its primary aim to discover transiting terrestrial planets. Exoplanets found with the SPECULOOS survey, like TRAPPIST-1 found by the SPECULOOS prototype survey \citep{gillon2017seven}, will enable the unique opportunity to observe their atmospheres for potential biological signals with future large observatories. However, to maximise the probability of finding such planets, one must minimise the red noise the survey is subjected to, including the noise induced by atmospheric PWV variability.

In this work, we have developed methods to photometrically correct for PWV-induced variability on differential light curves. Our study has been evaluated on photometric data from SPECULOOS' Southern Observatory (SSO), which consists of four 1~m class telescopes located at the Paranal Observatory \citep{jehin2018speculoos}. We utilised the standard observing modes of the LHATPRO instrument (located $\sim$200~m above and a 1.8~km lateral distance away from SSO), in addition to ground relative humidity and temperature measurements. These measurements have been used to estimate the PWV experienced by SSO observations by including an altitude difference correction and line-of-sight estimate. We have also assessed the impact of photometric contamination by the temporally-varying atmospheric PWV in several commonly used filter bandpasses, extending the work on the PWV correction described in \citet{murray2020photometry}. In the following sections, we describe our methodology, quantify the extent of PWV variability at Paranal, its effect on common red-visible~--~near-infrared bandpasses, and evaluate our correction method on photometric observations from SSO performed with the \textit{I+z'} bandpass. 

\section{Method of correction}\label{sec:theory}

The observed flux of an object through our atmosphere can be described as:

\begin{equation}\label{eq:pwv}
f(X, \textrm{PWV}, \textrm{T}_{\textrm{eff}}, t) = \int{W(\lambda,X,\textrm{PWV},t)\, R(\lambda)\, S(\lambda, \textrm{T}_{\textrm{eff}}, t)~\textrm{d}\lambda},
\end{equation}

\noindent
where $W(\lambda,X,PWV,t)$ is the atmospheric transmission as function of wavelength ($\lambda$), airmass ($X$), and PWV value with time ($t$). $R(\lambda)$ is the overall bandpass response as a function of wavelength. $S(\lambda, \textrm{T}_{\textrm{eff}}, t)$ is the flux density distribution of an observed object as a function of wavelength, effective temperature (T$_{\textrm{eff}}$), and time.

\begin{figure*}
  \includegraphics[width=\textwidth]{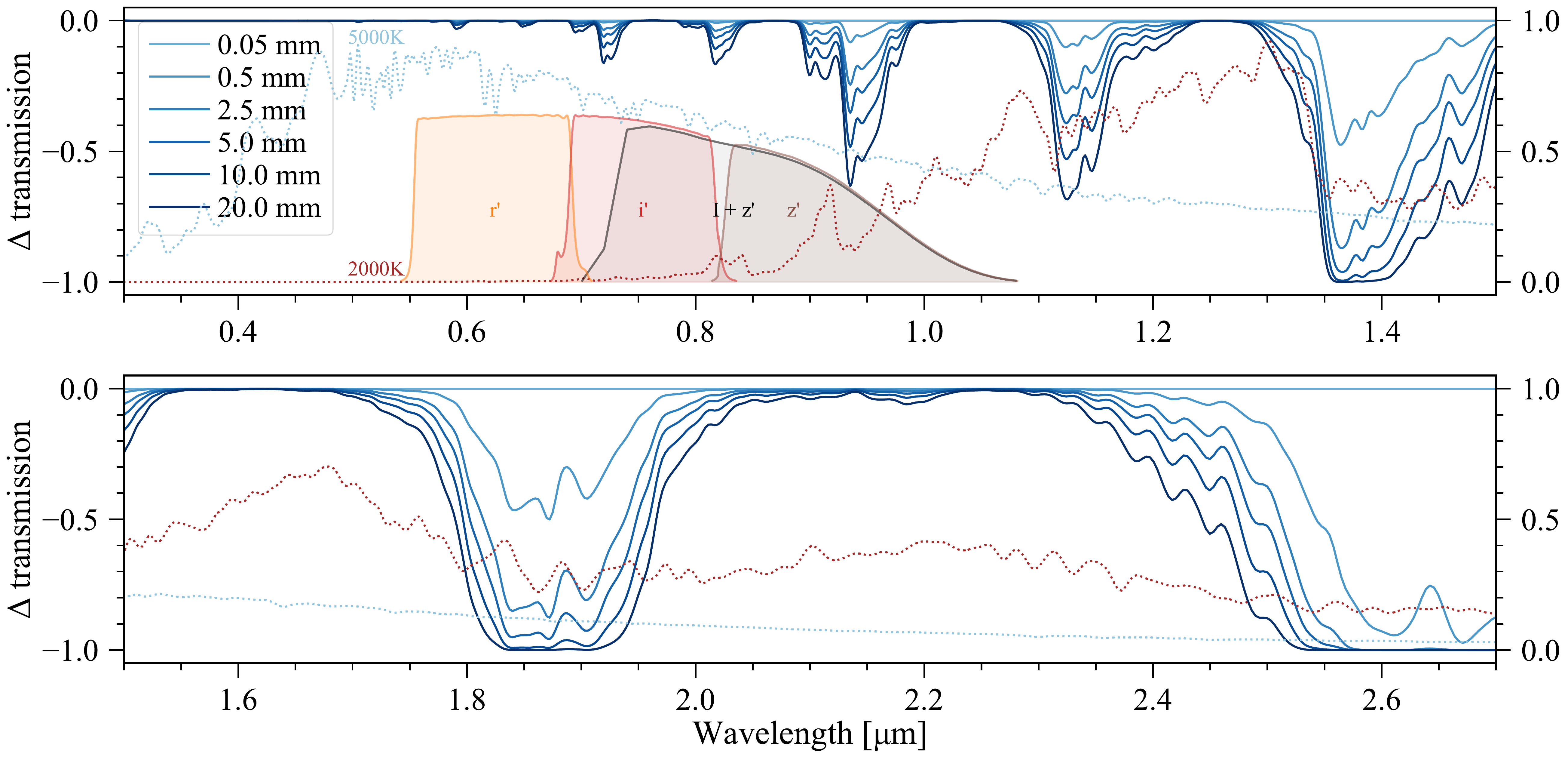}
  \caption{The fractional change of atmospheric transmission (left-hand axis) from the visible to near-infrared, at airmass 1, from a PWV of 0.05~mm to 20~mm is shown via a series of low resolution atmospheric spectra \citep[from the SkyCalc Sky Model Calculator, for 2400~m; ][]{Jones2013,Noll2012}. 5000~K and 2000~K stellar spectra are superimposed as dotted lines, from PHOENIX BT-Settl models \citep{2012RSPTA.370.2765A}. Four observational bandpasses (right-hand axis), \textit{r'}, \textit{i'}, \textit{z'}, \textit{I+z'}, with instrumental efficiencies of a telescope equipped with a deeply-depleted CCD accounted for are also shown.}
  \label{fig:h2ospectra}
\end{figure*} 

\begin{figure*}
  \includegraphics[width=\textwidth]{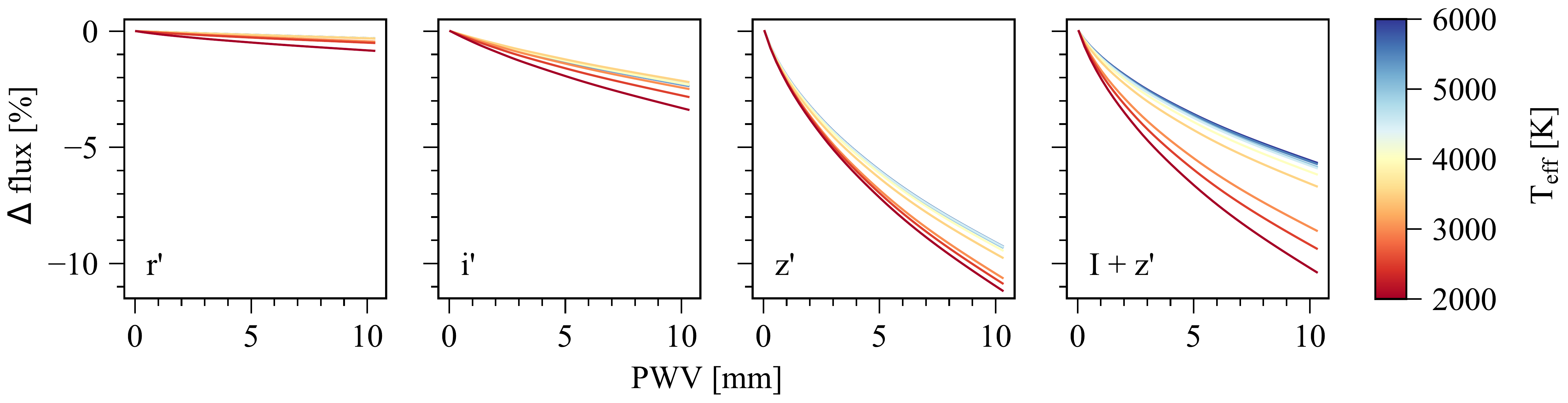}
  \caption{The change in flux as a function of PWV for different-temperature stars (from 6000~K to 2000~K in steps of 500~K) as modelled through four observational bandpasses, \textit{r'}, \textit{i'}, \textit{z'}, and \textit{I+z'} (profiles shown in Figure~\ref{fig:h2ospectra}), with respect to a 0.05~mm PWV atmosphere at airmass 1.}
  \label{fig:diff-flux}
\end{figure*}

In Figure~\ref{fig:h2ospectra}, the amount with which different PWV values affect visible and near-infrared atmospheric transmission is shown.  When Equation~\ref{eq:pwv} is applied, the change of flux as a function of PWV can be observed to differ significantly as a function of effective temperature; this is illustrated in Figure~\ref{fig:diff-flux} for the main bandpasses used by SPECULOOS (\textit{r'}, \textit{i'}, \textit{z'}, and \textit{I+z'}). For differential photometry, a second-order differential effect is thus induced when PWV is time-variable. 

One can divide out this effect by modelling the expected differential light curve produced by PWV variability, as per Equation~\ref{eq:pwv}. The difficulty in this approach stems from acquiring line-of-sight PWV values, in addition to having representative flux density distributions of the objects observed.

\subsection{The PWV grid}

For each filter bandpass we have developed a grid which outputs a time-independent value from Equation~\ref{eq:pwv}, when fed in observational parameters. These parameters are airmass (between 1~--~3), effective temperature (between 2000~K~--~36500~K), and PWV (between 0.05~mm~--~30~mm).

Each grid was built using atmospheric transmission profiles (0.5~{\textmu}m~--~2.0~{\textmu}m) from the online SkyCalc Sky Model Calculator by ESO \citep{Jones2013,Noll2012}, for 2400~m (the closest available altitude to SPECULOOS' site, using the `Entire year' `Entire night' profiles). In addition to these profiles, PHOENIX BT-Settl stellar models \citep{2012RSPTA.370.2765A} provided by the Python Astrolib PySynphot Python package \citep{laidler2008pysynphot} were used, with 2000~K as the lowest available temperature in the package. To build a stellar spectrum, one requires three parameters: effective temperature, metallicity, [Fe/H], and surface gravity, log(g). A representative set of stellar models were built using the parameters from \citet{2013ApJS..208....9P}\footnote{Specifically, the updated values from \url{https://www.pas.rochester.edu/~emamajek/EEM_dwarf_UBVIJHK_colors_Teff.txt} version 2019.3.22, where the 1960~K effective temperature was rounded to 2000~K.}, assigning a metallicity index of 0 to each spectra.

We generated 273 atmospheric profiles, a permutation of airmasses between 1~--~3 at 0.1 intervals, and PWV values between 0.05~mm~--~30~mm ([0.05, 0.1, 0.25, 0.5, 1.0, 1.5, 2.5, 3.5, 5.0, 7.5, 10.0, 20.0, 30.0]~mm). We then included a set of 91 stellar spectra to produce a grid of 24,843 $f(X, \textrm{PWV}, \textrm{T}_{\textrm{eff}}, t)$ values from Equation~\ref{eq:pwv} to interpolate in between. We have made the PWV grid code, \textsf{Umbrella}, publicly accessible on GitHub\footnote{\url{https://github.com/ppp-one/umbrella}}.

\subsection{PWV measurements -- estimating line-of-sight PWV}\label{sec:los_method}

We estimated line-of-sight PWV values by linearly interpolating over airmass between two measurements provided by the LHATPRO, one at zenith and another at an airmass of 2 (altitude of 30$^\circ$) from its cone scan measurements. Cone scan measurements were an average of 4 measurements at different bearings at a fixed altitude of 30$^\circ$, with its value scaled by the LHATPRO service as if observed at zenith.  Zenith and cone scan values were measured at $\sim$2~minute and $\sim$15~minute cadences respectively.  The LHATPRO also produces an all-sky scan value every $\sim$6~hours, but these were not considered here.

LHATPRO PWV measurements, for the period of this study, were acquired via ESO's Ambient Query Form\footnote{\url{https://archive.eso.org/wdb/wdb/asm/lhatpro_paranal/form}} in two stages. The first stage was acquired before the online archive was updated on 2020 Aug 28, the second stage was acquired after the update. Before the update, the data downloaded included zenith, cone scan, and all-sky measurements. Unfortunately for these measurements, a running average over 5 measurements and a $\sim$1~minute smoothing time-average was done by the service prior to downloading from the online archive. As a result, the different measurement types were not labelled or regularly spaced to be easily differentiable. To differentiate between the observing modes we used a peak detection method. Large peaks spaced at $\sim$6~hours intervals (assumed to be all-sky measurements) were first removed, then peaks spaced at $\sim$15~minute intervals were registered as the average of 4 cone scan values and 1 zenith value. Thus, only cone scan measurements which were larger or equal to zenith measurements of this dataset were subsequently identified. 

The data acquired after the update only included zenith measurements, without the issue of the 5 point running average and the $\sim$1~minute time-average. The other measurement types were not made public at the time of acquisition. Thus, with the previously acquired data, and the new zenith values, we could estimate non-moving averaged cone scan measurements by using the new zenith measurements from around the same measurement period.

\subsection{Accounting for altitude difference}\label{sec:alt_diff_method}

The majority of water vapour resides close to the ground, with a scale height between 1 and 3~km \citep{2017sgvi.confE..22K}. The altitude difference of $\sim$200~m between the LHATPRO at the VLT platform and SSO (located at the lower altitude) will therefore introduce an additional amount of PWV affecting our observations.

We estimated the missing vertical column of water vapour by integrating over the altitude difference, $\Delta h$, the estimated change in the density of water vapour, $\rho$ (in kg/m$^{3}$), between the respective sites,

\begin{equation}\label{eq:pwv-density}
\textrm{PWV} = \int{\rho~\textrm{dh}} \approx \frac{1}{2} \Delta h\left(\rho_{vlt} + \rho_{sso}\right).
\end{equation}

\noindent
To estimate the PWV, a linear change in water vapour density was assumed between SSO, $\rho_{sso}$, and the VLT platform, $\rho_{vlt}$, where the LHATPRO is located. This yields a value in kg/m$^2$, equivalent to PWV in millimetres when liquid density of water is 1000~kg/m$^3$.

To estimate the density of water vapour \citep{sensirion2009}, one can use ambient temperature, $T$ in $^\circ$C, and relative humidity, $RH$, measurements from the respective sites:

\begin{equation}\label{eq:density}
\rho = 0.2167~RH~\frac{6.112~\exp{\left(\frac{17.62~T}{243.12 + T}\right)}}{273.15 + T}~f_w(P),
\end{equation}

\noindent
where $f_w(P)$ is ``water vapour enhancement factor'' as a function of pressure, $P$, in hPa:

\begin{equation}\label{eq:enhancefactor}
f_w(P) = 1.0016 + 3.15\times 10^{-6}~P - 0.074~P^{-1},
\end{equation}

\noindent
where pressure was assumed fixed over time for the respective altitudes.

We used existing temperature and humidity sensors from the respective sites to produce density estimates. At SSO, we used the temperature and relative humidity sensor (Sensirion SHT15) on board a Boltwood Cloud Sensor II, with an assumed accuracy of $\pm~1$~$^\circ$C and $\pm~4$\% on relative humidity.\footnote{Accuracy taken from the sensor's datasheet: \url{https://sensirion.com/us/products/catalog/SHT15/} for the low humidity conditions seen in Paranal, Chile.} At the VLT platform, values of temperature and relative humidity were measured by a VAISALA METEOrological station 2m above the platform, with a quoted accuracy of $\pm~0.2$~$^\circ$C and $\pm~1$\% respectively \citep{sandrock1999vlt}.

The altitude difference was determined using Google Maps Elevation service \citep{google_elevation}, with LHATPRO's position on the VLT platform returning an altitude of 2633~m and SSO an altitude of 2446~m. For comparison, the altitude of VISTA's platform from Google Maps Elevation service was found to agree with ESO's stated value within 2~m. GPS-derived altitude values are also available at SSO, yielding a median value of 2482~m. The neighbouring facility to SSO, NGTS, has a quoted altitude of 2440~m \citep{wheatley2018next}, which led us to disregard the GPS-derived value.  It was decided to use the altitude difference given by Google Maps Elevation service of 187~m, with an assumed error of $\pm~10$~m. The lateral distance between the LHATPRO and SSO of 1.8~km was ignored.

\section{Results and discussion}

The following sections detail the results of applying the PWV correction to differential light curves observed by SSO over the course of approximately one year (2019 Feb 17~--~2020 Jan 31). 

\subsection{PWV variability at Paranal}

We experienced a median zenith value of 2.3~mm, a median cone scan value of 2.9~mm (pre-scaled by the LHATPRO service, as if observed at zenith), and a median PWV value of 0.26~mm calculated from the altitude difference, throughout our nightly observations.

\begin{figure}
  \includegraphics[width=\columnwidth]{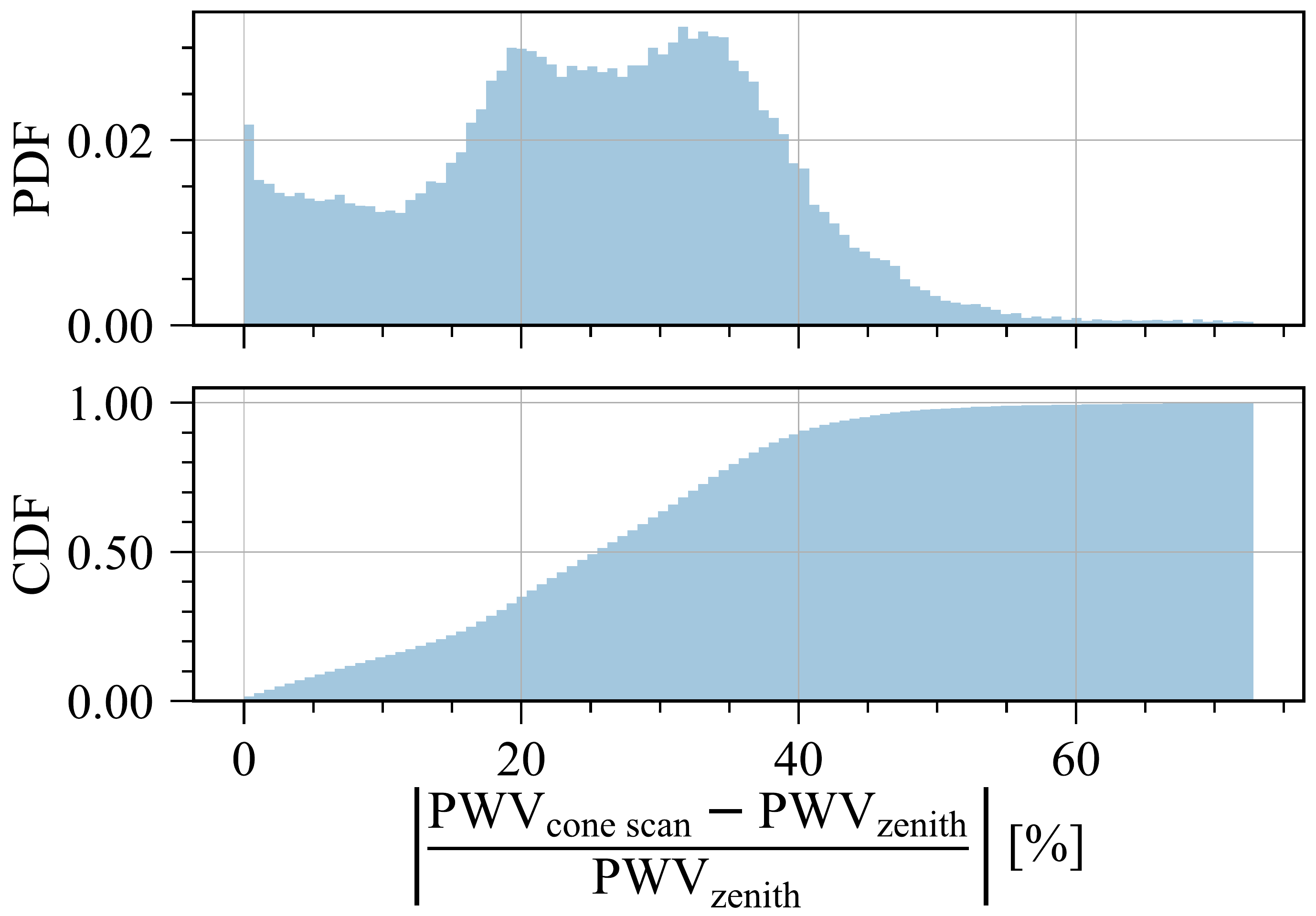}
  \caption{Probability density function (PDF) and cumulative distribution function (CDF) of the percentage difference between cone scan and zenith PWV measurements. Data from all publicly available nights, taking only dusk to dawn measurements.}
  \label{fig:pwv_con-zen/zen}
\end{figure} 

During our study, we assessed the absolute difference between 30$^\circ$ (airmass 2) cone scan values and zenith values, as shown in Figure~\ref{fig:pwv_con-zen/zen}. We found a median percentage difference of 26\%, as shown by the 0.5 CDF mark on Figure~\ref{fig:pwv_con-zen/zen}. However, since the detected cone scan PWV values were always to be higher than the zenith PWV values (due to the peak detection method described in Section~\ref{sec:los_method}), there is a possibility that the cone scan values are over reported in some instances. And as such, the PWV variation presented in Figure~\ref{fig:pwv_con-zen/zen} may be different in reality.

A similar difference was observed in further detail in \citet{Querel2014}, where they analysed 21 months of periodic all-sky scans, performed by the LHATPRO every 6~hours. From their study, they found a median all-sky variation of 10~--~26\% peak-to-valley, down to 27.5$^\circ$, with respect to the all-sky's zenith value. From this, they argued that zenith observations of PWV are sufficient for the general analysis and correction of astronomical data at Paranal. However, with the increasing interest in cooler stars, photometric surveys may rely on line-of-sight PWV observations, depending on their desired photometric accuracy.

\begin{figure}
  \includegraphics[width=\columnwidth]{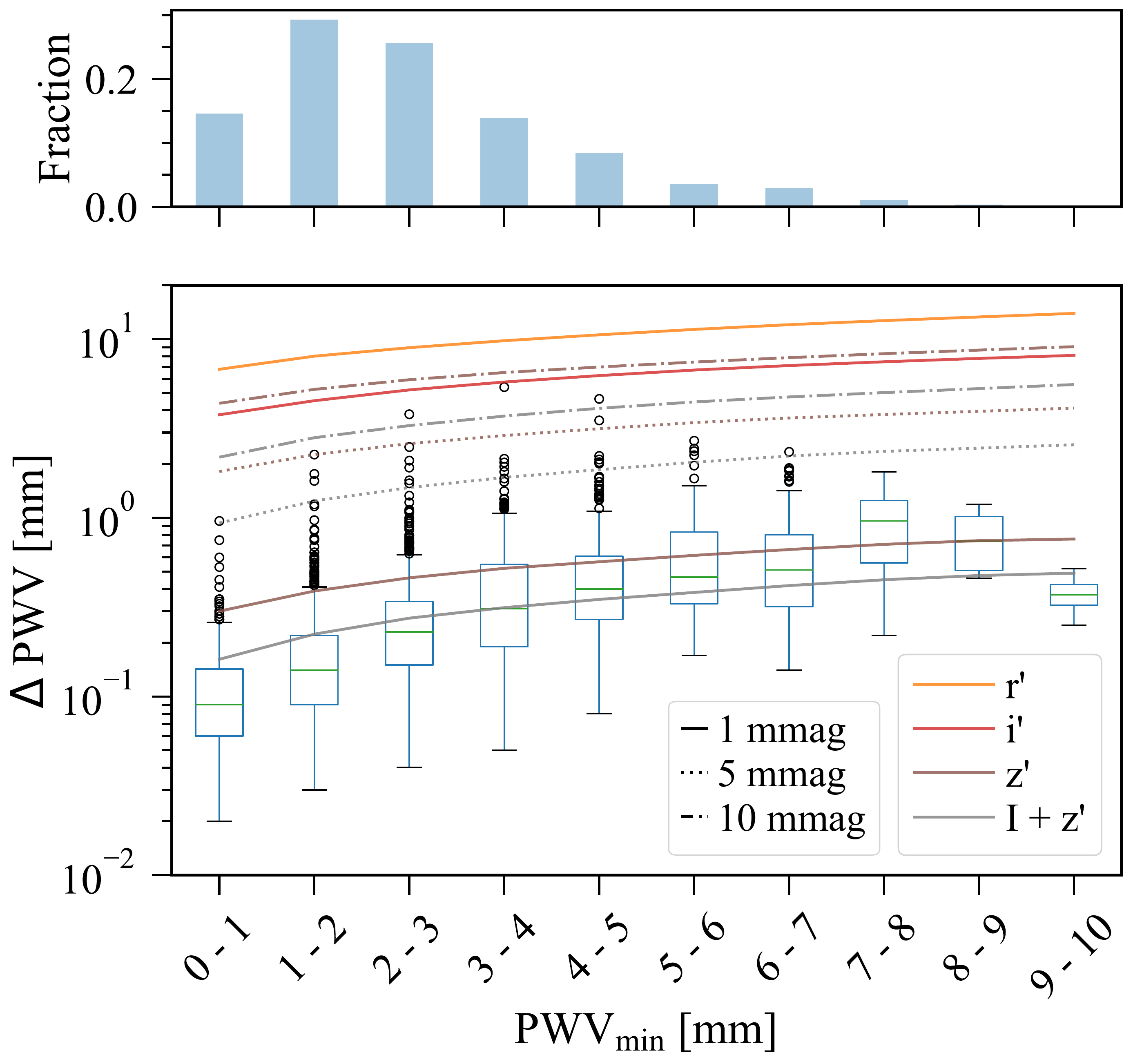}
  \caption{Sensitivity of common bandpasses and the \textit{I+z'} bandpass to PWV changes. Bottom: Measured PWV changes from evaluated consecutive blocks of one hour intervals of PWV zenith data (only between the hours of dusk to dawn). Here, we calculated the max-min change ($\Delta$PWV) of PWV within each evaluated hour. We grouped the measured changes into 1\,mm intervals of PWV$_\text{min}$, and plotted the respective box plots of the PWV changes from PWV$_\text{min}$ in each interval, where each box plot shows the standard median (green line), inter quartile range (IQR) of the lower (Q1) and upper (Q3) quartiles, and lower whisker (Q1 - 1.5*IQR) and upper whisker lines (Q3 + 1.5*IQR), with black circles denoting outliers of the whisker range. The coloured lines show the required $\Delta$PWV from the middle of each interval to induce a 1\,mmag, 5\,mmag, and 10\,mmag change in a light curve (when observing a 2700\,K target star and a 5200\,K comparison star), represented by the solid, dotted, and dash-dotted lines respectively for each \textit{r'}, \textit{i'}, \textit{z'}, and \textit{I+z'} bandpasses, in orange, red, brown and grey respectively. Changes at 5\,mmag and 10\,mmag for \textit{r'} and \textit{i'} were outside the model's PWV range of 30\,mm. Top: The fraction of hours contributing to the evaluated PWV interval ranges.}
  \label{fig:delta-pwv-1hour}
\end{figure} 

\begin{table}
    \centering
    \caption{Following Figure~\ref{fig:delta-pwv-1hour}, total proportion (in percent) of consecutive blocks of one hour of PWV zenith data, only between the hours of dusk to dawn, which display a maximum change to a light curve's flux ($\Delta$LC) in mmag, per bandpass.}
    \begin{tabular}{lrrrr}
        \hline
        $\Delta$LC [mmag] &     \textit{r'} &     \textit{i'} &     \textit{z'} &  \textit{I + z'} \\
        \hline
        0 -- 1         &  100.0 &  100.0 &  84.23 &   62.99 \\
        1 -- 2         &    0.0 &    0.0 &  11.72 &   26.10 \\
        2 -- 5         &    0.0 &    0.0 &   3.84 &    9.72 \\
        5 -- 10        &    0.0 &    0.0 &   0.20 &    1.06 \\
        > 10           &    0.0 &    0.0 &   0.00 &    0.12 \\
        \hline
    \end{tabular}
    \label{tab:delta-pwv-1hour-prob}
\end{table}

\begin{table}
    \centering
    \caption{Proportion (in percent) of transit-like structures over the evaluated dataset, within a range of depths in mmag, induced by PWV variability at zenith, per bandpass.}
    \begin{tabular}{lrrrr}
        \hline
        Depth [mmag] &     \textit{r'} &     \textit{i'} &     \textit{z'} &  \textit{I + z'} \\
        \hline
        0 -- 1  &  10.17 &  10.17 &  8.71 &    6.35 \\
        1 -- 2  &   0.00 &   0.00 &  0.98 &    2.77 \\
        2 -- 5  &   0.00 &   0.00 &  0.49 &    0.90 \\
        5 -- 10 &   0.00 &   0.00 &  0.00 &    0.16 \\
        > 10    &   0.00 &   0.00 &  0.00 &    0.00 \\
        \hline
    \end{tabular}
    \label{tab:delta-pwv-transit-prob}
\end{table}

Figure~\ref{fig:delta-pwv-1hour} and Table~\ref{tab:delta-pwv-1hour-prob} detail the extent to which common near-infrared bandpasses (for silicon based detectors) will induce second-order differential variability with PWV changes experienced in Paranal during this study. Here, the respective change in a light curve's flux, per bandpass, of observing a late M-dwarf (with a temperature of 2700~K and a 5200~K comparison star -- median temperatures of the sample in this paper) were simulated, using nightly PWV zenith data from the LHATPRO service. The \textit{I+z'} bandpass was found to be the most sensitive to PWV variability, followed by \textit{z'}, \textit{i'}, and \textit{r'}. On the hourly scale, the approximate expected transit duration for a temperate, rocky planet around a late-M or L-type star \citep{2008ASPC..398..475T}, the majority of PWV variability was found to induce a 0~--~1~mmag change in a light curve, within the typical photometric precision achieved by SSO. However, with \textit{z'}, and more significantly with the \textit{I+z'} bandpass, changes greater than 1~mmag were observed at 16\% and 37\% occurrence rates respectively, mimicking transit structures from time to time.

The accuracy on $\Delta$PWV was calculated to be 0.042~mm, propagated from LHATPRO's stated single-measurement precision of 0.03~mm, assuming the accuracy of better than 0.1~mm of a single measurement to be systematic. For the \textit{I+z'} bandpass, this accuracy level limits the correction to 0.4~mmag when PWV$_\text{min}$ is at 0.05~mm (the lower limit of the atmospheric models used), 0.2~mmag at Paranal's median value of 2.3~mm, and 0.1~mmag at a rarely seen 10~mm. This therefore suggests PWV measurements from the LHATPRO PWV measurement are sufficient for correcting sub-mmag-level changes in our \textit{I+z'} light curves, and even more so in the other band passes considered here.

The proportion of transit-like structures induced by zenith PWV variability  over the evaluated dataset was estimated in Table~\ref{tab:delta-pwv-transit-prob}, using the same data as Table~\ref{tab:delta-pwv-1hour-prob}. This was calculated by finding the proportion of consecutive hours that displayed a dip in one hour and followed an equivalent rise, within $\pm 25\%$, in the following hour. This method may miss some structures that occurred within an evaluated hour. Nonetheless, it suggests the occurrence rate of a significant transit-like structure, greater than $\sim$5~mmag, will not occur for the \textit{r'} and \textit{i'} bandpasses, and is a rare occurrence for both the \textit{z'} and \textit{I+z'} bandpasses.  The \textit{I+z'} bandpass will display a greater depth, given it has approximately twice the sensitivity to PWV changes than the \textit{z'} bandpass when observing late M and L-type stars. If such an event occurs, it is likely to be visible in co-current observations of similar temperature targets by other telescopes on site in the same bandpass, if they exist.

SSO is subject to additional PWV-induced effects due to the altitude difference and observing through a variety of airmasses. The additional previously unaccounted for PWV from the altitude difference between SSO and the LHATPRO, calculated with the method described in Section~\ref{sec:alt_diff_method}, was seen to generally scale with zenith PWV, however was not seen to have a direct relationship with zenith values. The derived altitude difference PWV values were dependent on the accuracies of the respective temperature and humidity sensors, and the accuracy of the altitude difference in meters. Propagating the quantified errors on the derived altitude difference PWV dataset provided a median fractional error of 18\%. The fractional error is likely higher in reality due to the approximation made in Equation~\ref{eq:pwv-density}.  The effect of the altitude derived PWV is evaluated in the following sub-section.

Similarly, the validity of linearly interpolating cone scan values over time, as described in Section~\ref{sec:los_method}, from a 15~minute to a 2~minutes time base (to match the cadence of zenith measurements), was assumed to be acceptable. To test this assumption, we used the $\sim$2 minute cadence zenith values, and every 7$^\text{th}$ value from the dataset linearly interpolated back onto the same time base as the original zenith observations, to simulate the cadence of the cone scan measurements. The difference between the values were found to be within $\pm$2\%, under 0.1~mm in most cases. The validity of interpolating between zenith and cone scan values over airmass is addressed in the following sub-section.

\subsection{Evaluation of the PWV correction}

We used observations made by SSO with its primary bandpass, \textit{I+z'}. SSO, with its four telescopes, made 1193 unique observations (divided by telescope/target/night) of 103 targets observed between 2019 Feb 17 to 2020 Jan 31 with the \textit{I+z'} bandpass. This amounted to a sum of 5420 hours of on-sky data. Differential light curves of these observations were produced with the SSO pipeline described in \citet{murray2020photometry}. 

For the correction, the target star was assigned an effective temperature derived in \citet{2020arXiv201102069S}, with an assumed systematic error of $\pm$100~K. The range of target temperatures evaluated was 2000~--~3000~K, with a median temperature of 2700~K. The comparison light curve, behaving as an artificial star, had an effective temperature assigned by the SSO pipeline -- a weighted sum of effective temperatures from the Gaia DR2 \citep{2018A&A...616A...1G} catalogue, of the stars used in the field. The range of  temperatures for the artificial star evaluated was 4000~--~6000~K, with a median temperature of 5200~K. For the \textit{I+z'} bandpass, the second-order effect induced by the artificial star behaves very similarly for any temperature above 4000~K, as illustrated in Figure~\ref{fig:diff-flux}. The target star was often the coolest star in the field, and as such, forming an ideal artificial star with a temperature equivalent to the target star was not  possible. Consequently, for the fields evaluated, this disparity in temperature between the target star and the effective comparison star would always yield a PWV induced effect under changing atmospheric conditions. One could potentially further improve the correction by fine-tuning the assigned effective temperature of the target, or by using the real spectra of the target -- this was not considered here.

When searching for transits, we seek to minimise any atmospherically induced variability on differentially resolved light curves. This is to maximise the likelihood of detecting real transit events. In this context, the observed dataset was evaluated in two regimes -- with a low and high pass temporal filter, with a dividing period of 120~minutes. The low pass temporal filter maintained variability that was greater than 120~minutes, and high pass filter maintained the variability shorter than 120~minutes. This was to demonstrate the correction's effect around and below the transit timescale, and on timescales where long-term variability may be induced. Table~\ref{tab:rms_change} details the effectiveness of the correction. Here, we evaluated the percentage change in the root-mean-square (RMS) on global light curves (30~minute binned) observed by SSO before and after the correction for the low and high pass temporally filtered domains. The correction was calculated with combinations of each of the PWV derived values (zenith, estimated line-of-sight, and altitude difference).

\begin{table*}
    \centering
    \caption{Evaluation of the RMS percentage change, $((\sigma_\text{LC~corrected} - \sigma_\text{LC})$/$\sigma_\text{LC}) \times 100$, of 103 global light curves (LC) with low and high pass filtering at 120~minutes. Showing the [10,25,50,75,90]th percentiles using the respective PWV derived values for correction, from zenith, zenith + altitude difference ($\Delta$Alt), estimated line-of-sight (Est. LoS), and Est. LoS + $\Delta$Alt.}
    \begin{tabular}{lrrrrrrrr}
\hline
{} &         \multicolumn{4}{c}{Low pass [\%]}                                          &          \multicolumn{4}{c}{High pass [\%]}                             \\
Percentiles [\%] &  Zenith &  Zenith + $\Delta$Alt &  Est. LoS &  Est. LoS + $\Delta$Alt &  Zenith &  Zenith + $\Delta$Alt &  Est. LoS &  Est. LoS + $\Delta$Alt \\
\hline
10          &    -0.2 &                 -0.2 &          -0.5 &                       -0.1 &      1.7 &                   1.7 &            4.3 &                         4.0 \\
25          &   -14.4 &                -14.4 &         -14.3 &                      -16.0 &      0.2 &                   0.2 &            0.4 &                         0.5 \\
50          &   -51.8 &                -52.9 &         -52.7 &                      -53.8 &     -1.2 &                  -1.3 &           -1.2 &                        -1.1 \\
75          &   -63.8 &                -64.5 &         -65.6 &                      -66.2 &     -4.3 &                  -4.3 &           -4.0 &                        -3.9 \\
90          &   -71.9 &                -72.1 &         -73.0 &                      -73.5 &     -9.1 &                  -8.6 &           -6.6 &                        -6.1 \\
\hline
\end{tabular}
    \label{tab:rms_change}
\end{table*} 

On timescales longer than 120~minutes, a zenith PWV based correction demonstrated a large median RMS percentage change of $-51.8$\%, when compared to an uncorrected differential light curve. This large percentage change in RMS is attributed to the long-term multi-millimetre variability of PWV, and as such demonstrates the correction's importance for variability studies on late-M and L-type stars, when using a bandpass like \textit{I+z'}. The addition of the altitude difference derived PWV to both zenith and estimated line-of-sight PWV made a marginal improvement, at $-52.9$\% and $-53.8$\% respectively. Using the estimated line-of-sight PWV alone yielded $-52.7$\%.

Whilst the estimated line-of-sight PWV presented a marginally better median percentage change, the assumption of circular symmetry around zenith in some instances may introduce false variability post-correction. To yield a more accurate correction, a new measurement mode for the LHATPRO is subsequently suggested: a continuous monitoring mode which constructs a low resolution all-sky map, maintaining a PWV accuracy of 0.1~mm, at a cadence better than 30~minutes (half the expected transit duration around a late M, L-type star, to follow Nyquist sampling).  Since we only seek to minimise long-term variability and false transit features, the current 2~minute cadence for zenith is faster than we currently need.

\begin{figure}
  \includegraphics[width=\columnwidth]{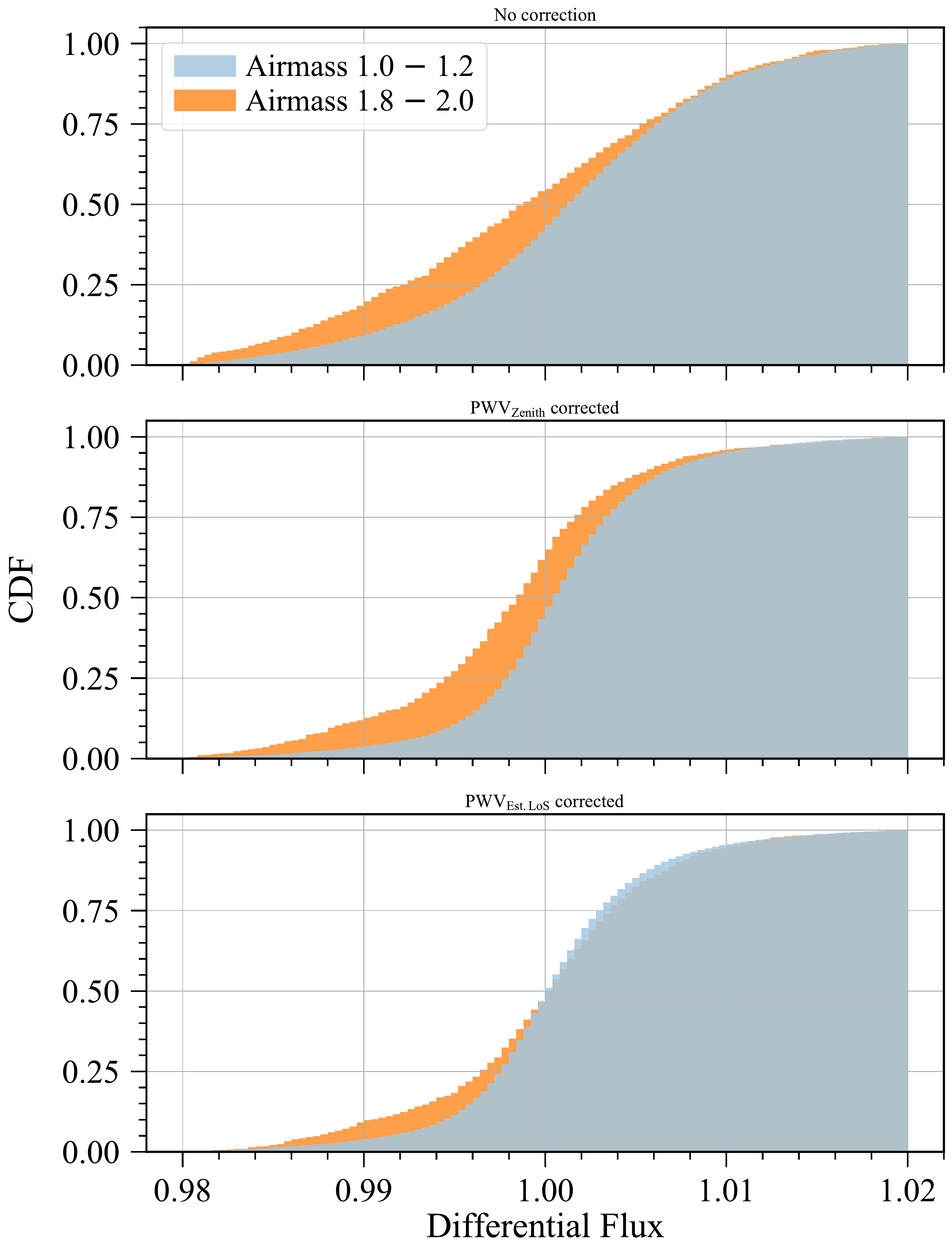}
  \caption{Assessing the correction's effect on differential flux at different airmass ranges. Cumulative distribution function (CDF) histograms of differential flux between 0.98 and 1.02 (of 0.005\,JD binned on-sky data) observed with the \textit{I+z'} bandpass. Global dataset assessed for when observations were in two different ranges of airmasses 1.0~--~1.2 and 1.8~--~2.0, with different levels of PWV derived corrections applied. First without any correction (top plot), zenith PWV correction (middle plot), and then estimated line-of-sight (Est. LoS) PWV correction (bottom plot).}
  \label{fig:airmass}
\end{figure}

The estimated line-of-sight PWV also minimised a previously uncorrected airmass effect: an observed decrease of differential flux when transitioning to higher airmasses. This effect is illustrated in Figure~\ref{fig:airmass}. Without the correction, a $-3.0$\,mmag differential flux change from the median differential flux at airmasses between 1.0 and 1.2 to the median differential flux at airmasses between 1.8 and 2.0 was observed. This change improves slightly with the correction derived with zenith PWV, and noticeably improves with the estimated line-of-site PWV with a median differential flux increase of $0.1$\,mmag observed from airmasses between 1.0 and 1.2 to airmasses between 1.8 and 2.0, versus the $-3.0$\,mmag observed without line-of-sight correction. We performed a two-sample Kolmogorov–Smirnov test to assess the statistical significance of this result, where we assessed the cumulative distributions of differential flux at high airmasses (1.8~--~2.0) corrected with zenith PWV values (Figure~\ref{fig:airmass}, middle plot, orange CDF), $F(x)$, to the correction derived from estimated line-of-site PWV values (Figure~\ref{fig:airmass}, bottom plot, orange CDF), $G(x)$. Our null hypothesis being that $F(x) \geq G(x)$ for all airmasses. The resulting Kolmogorov–Smirnov test yielded a p-value of 0.992, leading us to accept the null hypothesis, confirming our observation that line-of-sight PWV data helps to minimise the decreasing of differential flux at higher airmasses. From the distributions seen in Figure~\ref{fig:airmass}, we can also see the reduced spread of differential flux when the correction is applied, as quantified in Table~\ref{tab:rms_change}.

On timescales shorter than 120~minutes, a median RMS percentage change of around $-1$\% for all PWV derived values was observed. However, without the ground-truth of stellar variability at this scale, it's difficult to argue if one PWV derived correction is better than another when solely based on the RMS percentage change. Whilst it has been seen to correct transit-like features in \citet{murray2020photometry} and one example in the next sub-section, the correction at shorter time scales was seen to increase the RMS in about a quarter of the instances evaluated. Once again, without the ground truth at this scale, or extended trend modelling, it's currently difficult to validate the true extent of improvement at this scale. It is therefore recommended to manually vet where possible. Nonetheless, with the dataset evaluated here, we have not seen instances where the correction has induced significant false variability. The cases where this may occur would be likely due to PWV measurement failure, or significant line-of-sight differences from observation and PWV measurement.

\subsection{PWV correction in action}

\begin{figure}
  \includegraphics[width=\columnwidth]{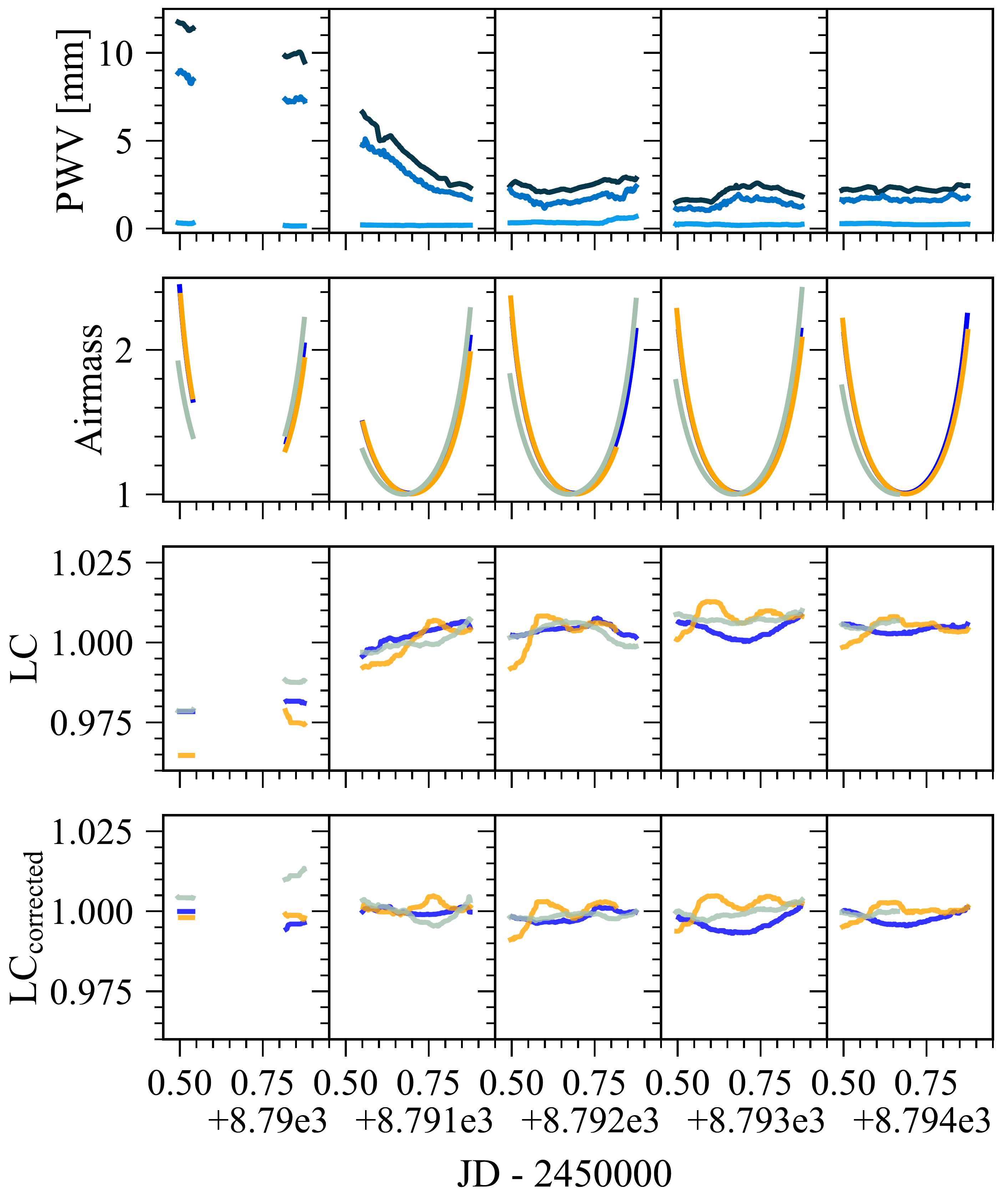}
  \caption{Five consecutive nights observed by three SSO telescopes of three different targets, one per telescope - distinguished by the different colours in the lower three plots. The first night is partially missing data due to bad weather. First row: PWV values from cone scan (dark blue), zenith (blue), altitude measurements (light blue) are shown. Second row: Airmass of the respective observations, which shows the transition between zenith and the airmass cone scan PWV measurements are made (airmass 2), used for the line-of-sight PWV estimate. Third row: Differential light curves (LC) without PWV correction. Bottom row: Differential light curves with PWV correction, using estimated line-of-sight + altitude difference PWV.  The global light curves were normalised over the assessed period. The effective temperatures of the target/comparison stars were 2500/5000, in orange; 2600/4700, in light green; 2700/4300, in blue, respectively. The light curves shown are with a 120~minute low pass filter applied -- trends from 120\,minute windows using a median method. Here, the average 120\,minute binned error was sub-mmag for all light curves.}
  \label{fig:global_eg}
\end{figure}

In this sub-section, examples of the correction with the \textit{I+z'} bandpass are shown, on the global and nightly scale. The first example, in Figure~\ref{fig:global_eg}, shows five consecutive nights of observations of three targets of similar temperature (2500~-~2700~K), with the global light curves normalised over the assessed period and with a 120~minute low pass filter applied. Between the first and second night, a large PWV change was observed. The resulting second-order effect is evident in the uncorrected light curves, showing around a 30~mmag change for all the targets. The behaviour on shorter timescales, however, was not always comparable between targets due to the respective variability that is often seen with M dwarfs \citep{gunther2022complex}. Beyond the second night on Figure~\ref{fig:global_eg}, the PWV changes were less significant on the light curves by eye. However, a level of difference is still visible. No other systematic effects were seen to correlate with the observations.

The respective PWV values from zenith and cone scan followed a similar trend for the five nights. A similar PWV zenith and cone scan behaviour was seen for the remaining dataset. The altitude difference derived PWV stayed relatively constant with the exception of the third night, where a small increase was observed at the end of the night, with an opposing change seen at zenith around the same time. A layer of water vapour likely transitioned from above the LHATPRO to the layers below it. A sharp dip in flux in all the light curves was observed at the same time. The correction removed the majority of the structure, however some residual in the shape was left in all the light curves, suggesting some amount of water vapour was unaccounted for in the line-of-sight and altitude estimates. The residual shape could likewise be attributed to inaccuracies in the target and comparison stars temperature estimates, where a higher target temperature estimate would have under corrected the variability. This could be similarly argued for the light green target between the first and second night. Co-current observations of this target could not be found to rule out stellar variability. 

\begin{figure*}
  \includegraphics[width=\textwidth]{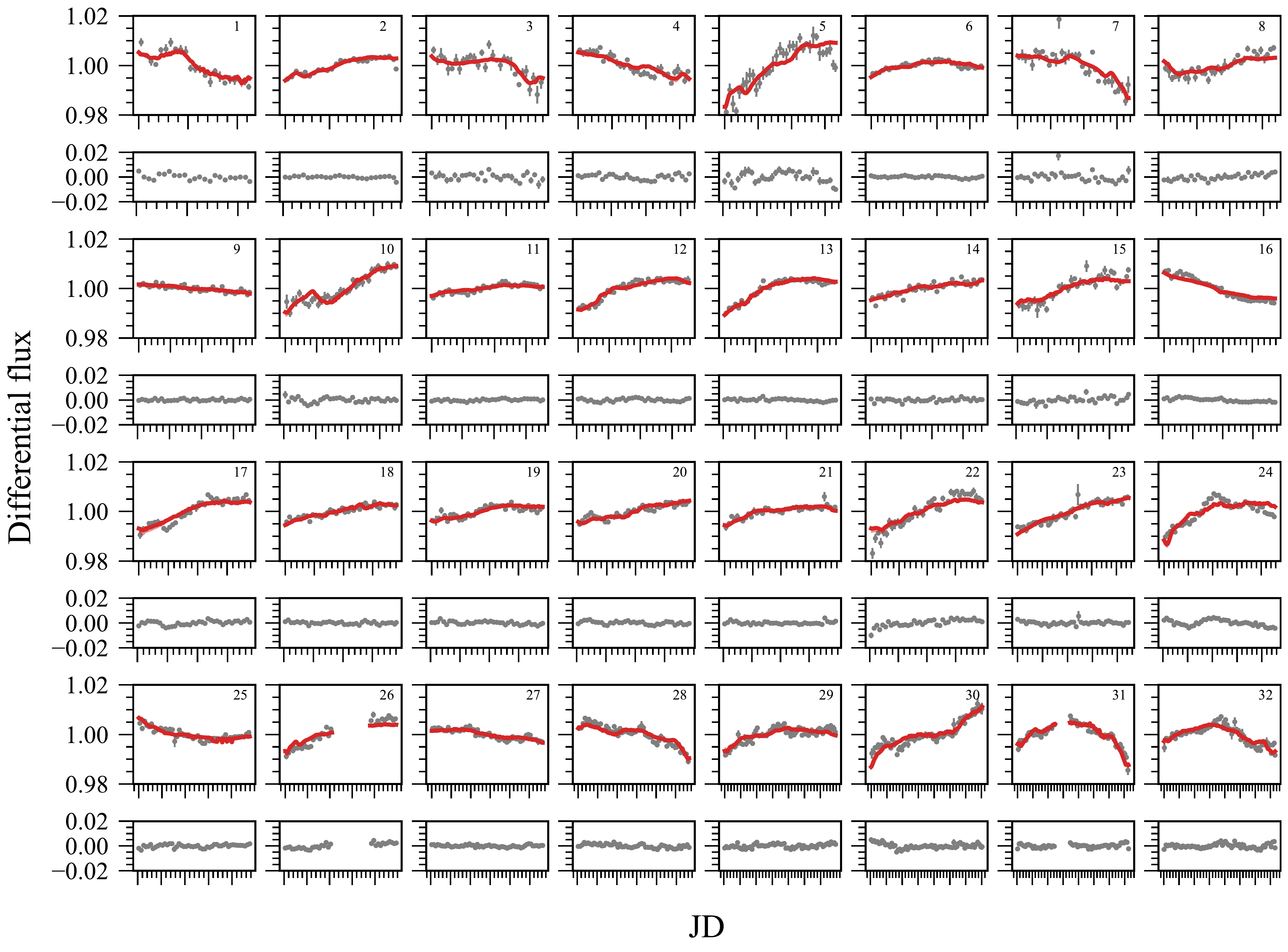}
  \caption{A selection of 32 observations of 19 different targets observed with the \textit{I+z'} bandpass which matched closely with the expected trend from the correction. Top row: 0.005 JD binned (7.2 minutes) uncorrected differential flux light curves in grey. Expected trend from the correction in red, using estimated line-of-sight + altitude difference PWV and knowledge of the target and comparison stars effective temperatures. In shaded red, although the effect is not visible for the majority of examples, the expected trend from the correction also plotted using $\pm$100\,K from the target's effective temperature. Second row: Residual of the correction (observed data - expected trend) of the above subplot. Row order then repeats. Ordered from shortest to longest timescales, where the major ticks on the x-axis are 0.05~JD (72~minutes).}
  \label{fig:pwv-match}
\end{figure*} 

If one were to adopt the methodology in \citet{irwin2011angular}, briefly described in Section~\ref{sec:intro}, for timescales greater than 120~minutes at 30~minute bins, one would need at least 9 co-current observations of similar temperature objects to yield a common mode with variability at the sub-mmag scale. At timescales below 120~minutes, however, one would need an unrealistic 90 co-current observations. This was calculated by evaluating the overall RMS SSO experienced (with correction) at the respective scales, and assuming a RMS scaling of root sum the total number of co-current observations.

In Figure~\ref{fig:pwv-match}, 32 examples of single observations are shown. In these examples, the uncorrected differential light curves displayed a close resemblance to the expected trend modelled by the PWV grid, with PWV values from the estimated line-of-sight and altitude difference. These were found by finding the nights where the standard deviation of the night's data was significantly reduced by the correction. Such examples often occur on quiet targets, where the second-order effects are very evident on both short and long timescales. For example, on the shorter timescales, false transit features have been mostly corrected for in subplots labelled 1, 3 and 12. On the longer timescales, a range of other variabilities are closely matched, such as an inverse airmass like shape in the subplot labelled 31. There is an instance in Figure~\ref{fig:pwv-match} where a transit like feature is induced by the correction (subplot labelled 10). The exact origin of this feature is unknown, most likely a line-of-sight induced feature. Manual vetting is therefore recommended when such events occur. The remaining light curves matched closely the expected trend modelled by the PWV grid, with many examples beyond the 32 presented light curves which similarly match the modelled trend.

\section{Conclusions}

We have developed a method of modelling and mitigating the second-order effect induced by PWV on time-series photometric data.  This has been enabled by leveraging the accurate measurement modes provided by an onsite radiometer, the LHATPRO, and local environmental sensor data. The developed tool, the PWV grid, has proven to be an essential for SSO, and we believe it can help other studies who are likewise sensitive to PWV and have access to accurate PWV data. The PWV grid code, \textsf{Umbrella}, has been open-sourced on GitHub\footnote{\url{https://github.com/ppp-one/umbrella}}.

We found, for removing transit-like structures and long term variability on late M- and early L-type stars, the LHATPRO's single measurement PWV accuracy of better than 0.1~mm, and precision of 0.03~mm, is sufficient to eliminate sub-mmag level PWV induced photometric effects for the \textit{I+z'} and \textit{z'} bandpasses, and more than sufficient for the \textit{i'} and \textit{r'} bandpass. The \textit{I+z'} bandpass was shown to be exceptionally sensitive to second-order effects induced by PWV, and without aid of the correction, the bandpass significantly limits ones ability to do variability studies on late M- and L-type stars. On the transit timescale, the bandpass is sensitive to variability which may mimic transit-like structures on the rare occasion with Paranal's level of PWV variability.

PWV data from zenith was found to be sufficient to support the majority of the correction needed for the four telescopes at SSO. However, through our use of zenith and cone scan measurement modes, there are residual second-order effects induced at higher airmasses which would require line-of-sight measurements to accurately correct for. We have therefore recommended a continuous all-sky observing mode for the LHATPRO, such to support more accurate line-of-sight estimates for our multiple telescopes at Paranal.

The additional PWV derived from the altitude difference between the LHATPRO and SSO was accounted for through the use of local environment sensors, and was shown to improve the correction on timescales longer than 120~minutes. On shorter timescales, a more accurate method of accounting for the altitude difference may be needed. If one does not have access to PWV data, then optimising the bandpass for the survey is necessary.

\section*{Acknowledgements}

The research leading to these results has received funding from the European Research Council (ERC) under the FP/2007--2013 ERC grant agreement n$^{\circ}$ 336480, and under the H2020 ERC grants agreements n$^{\circ}$ 679030 \& 803193; and from an Actions de Recherche Concert\'{e}e (ARC) grant, financed by the Wallonia--Brussels Federation. We also received funding from the Science and Technology Facilities Council (STFC; grants n$^\circ$ ST/S00193X/1, ST/00305/1, and ST/W000385/1). This work was also partially supported by a grant from the Simons Foundation (PI Queloz, grant number 327127), as well as by the MERAC foundation (PI Triaud), and the Balzan Prize foundation (PI Gillon). PPP acknowledges funding by the Engineering and Physical Sciences Research Council Centre for Doctoral Training in Sensor Technologies and Applications (EP/L015889/1). ED acknowledges support from the innovation and research Horizon 2020 program in the context of the Marie Sklodowska-Curie subvention 945298. MNG acknowledges support from the European Space Agency (ESA) as an ESA Research Fellow. This publication benefits from the support of the French Community of Belgium in the context of the FRIA Doctoral Grant awarded to MT. LD is an F.R.S.-FNRS Postdoctoral Researcher.

We would like to thank Alain Smette for his help and guidance with the LHATPRO data.

\section*{Data Availability}
The data underlying this article is not currently public. The SPECULOOS-South Consortium will make all SPECULOOS-South Facility reduced data products available to the ESO Science Archive Facility following the regular Phase 3 process as described at \url{http://www.eso.org/sci/observing/phase3.html}.




\bibliographystyle{mnras}
\bibliography{paper} 








\bsp	
\label{lastpage}
\end{document}